# Random Matrix Theory and the Failure of Macro-economic Forecasts

Paul Ormerod (**Pormerod@volterra.co.uk**) *

and

Craig Mounfield (**Craig.Mounfield@volterra.co.uk**)

Volterra Consulting Ltd

The Old Power Station
121 Mortlake High Street
London SW14 8SN

18<sup>th</sup> January 2000

* Corresponding Author

**Abstract**

*By scientific standards, the accuracy of short-term economic forecasts has been poor, and shows no sign of improving over time. We form a delay matrix of time-series data on the overall rate of growth of the economy, with lags spanning the period over which any regularity of behaviour is postulated by economists to exist. We use methods of random matrix theory to analyse the correlation matrix of the delay matrix. This is done for annual data from 1871 to 1994 for 17 economies, and for post-war quarterly data for the US and the UK. The properties of the eigenvalues and eigenvectors of these correlation matrices are similar, though not identical, to those implied by random matrix theory. This suggests that the genuine information content in economic growth data is low, and so forecasting failure arises from inherent properties of the data.*

1.     Introduction

There is growing interest amongst physicists in economic systems (for example, [1] - [4]). Much of this is focused upon financial markets and movements in asset prices. Techniques reported in physics journals can also be helpful in understanding other aspects of economic activity. We report here how the theory of random matrices can inform us about the reasons for the persistent failures of prediction of the overall rate of growth of the economy.

We offer an explanation for the poor forecasting record of economists. The data contain insufficient information for predictions to be made which have any reasonable degree of systematic accuracy over time. This paper argues that it is not possible in the current state of scientific knowledge to improve the accuracy of macro-economic forecasts for the rate of growth of total output (denoting total output subsequently by GDP - Gross Domestic Product - as is the convention in economics[1]).

In the post-war decades economists began to construct empirical models of the economy at the aggregate level - 'macro-economic' models in the phraseology of economists [5]. The models are constructed using a combination of economic theory to specify the relationships, and of statistical regression analysis of data to place values on the parameters in the relationships.

An important use of such models is as an input to short-term forecasting of the economy. Factors such as inflation, unemployment, consumer spending and the overall level of output in the economy are covered by such forecasts. Predictions are made and published regularly by governments throughout the West, and private sector institutions, ranging from research institutes to Wall Street and City of London firms, also carry out such forecasts. The results of the forecasts are in turn used as inputs to significant policy decisions, such as setting interest rates and deciding monetary policy. Additionally commercial businesses are also influenced by these forecasts regarding the scale of their commercial operations.

---
[1] All references to GDP are to real GDP ie: GDP measured net of inflation

Over the past thirty years in particular, a track record of forecasts and their accuracy has been built up. Economists disagree about how the economy operates, and these disagreements are reflected in, amongst other things, the specification of the relationships in macro-economic models. But, over time, no single approach has a better forecasting record than any other. Indeed, by scientific standards, the forecasting record is very poor, and a major recent survey of macro-economic forecasting [6] concludes that there is no real evidence which suggests that accuracy is improving over time.

As examples of the one-year ahead forecasting record for GDP growth, for the US economy recessions have not generally been forecast prior to their occurrence, and the recessions following the 1974 and 1981 peaks in the level of output were not recognised even as they took place[2] [6]. Further, growth has generally been overestimated during slowdowns and recessions whilst underestimates occurred during recoveries and booms [7]. For the UK, the predictions of the Treasury over the 1971-1996 period have been at least as good as those of other forecasters, but the mean absolute annual forecast error for these one-year ahead predictions was 1.45% of GDP, compared to an actual mean absolute change of 2.10% [8]. In 13 European countries over the 1971-1995 period, the average absolute error was 1.43% of GDP, compared to the average annual change of 2.91% [9].

In general, the forecasting record exhibits a certain degree of accuracy in that the average error over time is smaller than the size of the variable being predicted. But the error is still large compared to the actual data, and most of the accurate forecasts were made when economic conditions were relatively stable [6].

2. **Analysis of Macroeconomic Data**

---

[2] Economic data, except in financial markets, does not appear immediately, and it can be several months before a preliminary estimate of the level of output in a given period becomes available.

Recent work by physicists [1] – [4] draws on the theoretical properties of random matrices to examine the behaviour of asset price returns in financial markets. They note standard results on the density of the eigenvalues of the correlation matrix of a random matrix and compare these with the eigenvalues of empirical correlation matrices of asset price returns. For an T x N random matrix, the range of the eigenvalues of its correlation matrix is given in the limit when Q is infinitely large by [10]:

$$\lambda_{max,min} = \sigma^2 ( 1 \pm 1/ Q^{0.5} )^2 \qquad (1)$$

where $\sigma^2$ is equal to the variance of the elements of the correlation matrix, and Q = T/N.

In financial markets, data is available over time for a large number of individual series. In contrast, for each country, such as the US, say, there is only one series available over time for the rate of growth of GDP. In such circumstances, when just a single historical series exists, a standard method in the time series analysis of dynamic systems is to form a delay matrix from the original series [11].

Let **x**(t) be a T x 1 vector of observations of the rate of growth of GDP at time t, where t runs from 1 to T. We form a delay matrix, **Z**, such that the first column of **Z** is **x**(t), the second **x**(t-1), and so on through to **x**(t-m) in the (m+1)th column.

By suitable choice of m, the delay matrix can span what is usually thought of as the time period of the business cycle in economics, in other words the period over which any regularity of behaviour of the growth of GDP might be postulated to exist. Individual cycles vary both in terms of amplitude and period. A major study [12] many years ago concluded that the period ranged from some two to twelve years, a range which still commands broad assent amongst economists, though the upper bound might now be felt to be slightly high. A recent result [13] reports a power spectrum of post-war US quarterly GDP growth which is concentrated at frequencies between 2.33 and 7 years, although the concentration is very weak.

The properties of the correlation matrix **C** of this delay matrix where

$$\mathbf{C} = 1/(T - m)\, \mathbf{Z}^T \mathbf{Z}$$

may be compared to a 'null hypothesis' purely random matrix which one could obtain from a finite time series of strictly uncorrelated series. Deviations of the properties of **C** from the correlation matrix of the random series would suggest the existence of true information.

Specifically, we examine the eigenvalues of the empirical correlation matrix, **C**, and compare them to the range of eigenvalues of the correlation matrix of a purely random matrix, given by (1). If the computed eigenvalues lie within the theoretical limits of those of a random matrix, we can infer that any correlations which exist between the rate of growth of GDP at time t and any lagged values of itself are essentially random and have no real significance. In other words, the rate of growth of GDP at time (t+1) cannot be systematically predicted with any real degree of accuracy using information contained in past rates of growth of GDP.

Carefully constructed annual data on GDP per capita in 17 advanced capitalist economies from 1870 to 1994 is provided in [14][3]. These include the US, the UK, Germany, France, Italy and Japan. For each of these economies, after calculating the annual percentage rate of growth and taking lags up to 12 years, we obtain a delay matrix of dimension 112 x 13.

For a random matrix of this dimension, (1) indicates that the eigenvalues of its correlation matrix should fall in the range 0.435 to 1.797 (where the variables in the matrix are independent identically-distributed normal random variables scaled to have constant unit volatility). However, (1) only holds in the limit, and so we examined the possible existence of small-sample bias. Computing the eigenvalues of the correlation

---

[3] For Japan, estimates are available from 1885, and for Switzerland from 1900

matrix of 5,000 such random matrices did in fact suggest a slight bias, with the range falling between 0.329 and 2.005.  The summary statistics for the 65,000 eigenvalues computed are as follows:

min 0.329; 1st quartile 0.727; mean 1; 3rd quartile 1.247; max 2.005

The eigenvalues of the correlation matrix of the delay matrix formed from the annual rate of growth of GDP can be compared with these theoretical and empirical ranges.  For no less than 11 out of the 17 countries, the eigenvalues all lie within the theoretical limits given by (1).  France and Denmark each have one eigenvalue which lies within the theoretical maximum value in the limit and the empirical maximum (1.803 and 1.828 respectively), and Canada (1.960 and 1.882) and New Zealand (1.955 and 1.887) each have two.

Only in the case of the US and the UK do we find eigenvalues outside the empirically calculated range of those for a random matrix.  Even here, they are not large, being 2.174 and 2.051 for the US and 2.181 and 2.055 for the UK.  A potential economic explanation of this finding is that over the 1871-1994 period, only the UK and, latterly, the US have been the dominant world economic power, and therefore able to operate, albeit to a very limited extent, as an isolated system.

As a comparator, we took a time series where it is known that accurate genuinely ex ante predictions can be obtained, namely the mean annual sunspot data.  This is shown, for example, with a low-dimensional non-linear autoregressive model [15].  We took data for the same length as that available for annual GDP growth, and calculated a delay matrix again using 12 lags.  The resulting largest eigenvalues were 5.391 and 4.126, far larger than the empirical maximum.  The third largest eigenvalue was also greater than the latter value, at 2.213.  The smallest eigenvalues were far below the minimum, being close to zero.  In other words, the technique shows that the correlations between lagged values of the sunspot series do contain genuine information.  This is consistent with the fact that the series can be predicted with systematic accuracy over time.

Figure 1 plots the density of the eigenvalues calculated for all 17 economies, and the density of the eigenvalues of the correlation matrices of 5,000 random matrices discussed above. The latter looks much smoother when plotted, for there are 65,000 such eigenvalues compared to only 221 for the economic data.

On a formal Kolmogorov-Smirnov goodness-of-fit test, the hypothesis that the cumulative density function of the economic data eigenvalues is not significantly different from that of the cumulative density function of the random data eigenvalues is rejected at a p value of 0.073. In other words, at the conventional level of $p = 0.05$, the null hypothesis is not rejected. However, the lack of rejection is not dramatic, and careful inspection of the economic data eigenvalues reveals that almost all countries have at least one eigenvalue close to the theoretical maximum for a random matrix - as the small peak at around 1.7 in Figure 1 shows. In other words, the results suggest a small amount of genuine information in the movements of GDP.

This suggestion is confirmed by examination of the eigenvectors. Random matrix theory predicts that the components of the normalised eigenvectors of a Gaussian orthogonal ensemble matrix will follow a normal distribution [1] – [4]. This null hypothesis is rejected on a Kolmogorov-Smirnov test even for $p = 0.05$ for 8 of the 17 countries, and is rejected in the range $0.05 < p < 0.10$ for a further 2.

Calculation of the inverse participation ratio of the eigenvectors also reveals a certain amount of information in the data. For eigenvector k the inverse participation ratio, $I_k$, is defined [1] – [4] as the sum of the fourth power of the components of the eigenvector.

An eigenvector with identical components has $I_k = 1/m$, and one with one component equal to 1 and rest zero has $I_k = 1$. Figure 2 plots $I_k$ against the eigenvalues for the United States.

In this case, $1/m = 0.077$, so the $I_k$ all lie somewhat above this value. For one of the largest eigenvalues, there is a small peak in $I_k$, and a rather larger one for some of the smallest eigenvalues. Following [3], we interpret this latter result as implying that the eigenvectors are to some extent localised with a small number of the lags contributing to them rather more than the others.

We also considered the quarterly rate of growth of GDP in the US and the UK. Data on the economy as a whole is not in general available at a higher sampling frequency (for example, monthly). Quarterly data is available only for the post-war period, in the case of the US from 1947Q1 through 1999Q3, and for the UK from 1955Q1 through 1999Q1.

The theoretical range of eigenvalues for the correlation matrix of the US data delay matrix - again using lags of up to 12 years - is from 0.209 to 2.380. A calculation using 5,000 random matrices of the same dimension produced a range of 0.154 to 2.541. The actual eigenvalues calculated from US data lie somewhat more decisively outside this range than those calculated from the longer run of annual data. The largest two are 3.32 and 3.16, and the third largest is 2.451. This suggests a certain degree of information in the data correlations, and hence in principle a certain degree of predictability. The Federal Reserve do appear to have had some success in actual prediction during the 1990s.

For the UK, the results with quarterly data suggest less information content in the post-war period than is the case with the longer run of annual data. The range of eigenvalues from random matrices of the same dimension as the UK delay matrix is from 0.118 to 2.551. The largest eigenvalue calculated from the UK data is, by coincidence, also 2.551. These findings for the US and UK in the post-war period are consistent with the hypothesis formulated above to account for the longer run results, namely the ability of the dominant world economic power to operate as a partially isolated system. In the post-war period, this has not been true of the UK, but certainly is the case for the US.

## 3. Conclusions

In summary, the poor forecasting record of GDP growth by economists appears to be due to inherent characteristics of the data, and cannot be improved substantially no matter what economic theory or statistical technique is used to generate them. Over what is thought of as the time period of the business cycle in economics, in other words the period over which any regularity of behaviour of the growth of GDP might be postulated to exist, the genuine information content of correlations over time in the data is low.

[1]     R. N. Mantegna and H. E. Stanley, *An Introduction to Econophysics,* Cambridge University Press (1999) and references therein

[2]     L. Laloux, P. Cizeau, J-.P. Bouchaud and M. Potters, 'Noise Dressing of Financial Correlation Matrices', Phys. Rev. Lett **83**, 1467 (1999)

[3]     V. Plerou, P. Gopikrishnan, B. Rosenow, L. A. Nunes Amaral and H. E. Stanley, 'Universal and Non-universal Properties of Cross Correlations in Financial Time Series', Phys. Rev. Lett **83**, 1471 (1999)

[4]     S. Drozdz, F. Grummer, F. Ruf and J. Speth, 'Dynamics of Competition Between Collectivity and Noise in the Stock Market', cond-mat/9911168

[5]     A very early example of this is Klein L.R. (1947), 'The use of econometric models as a guide to the policy process', *Econometrica*, vol.15,pp.111-152

[6]     Stekler, H. and Fildes, R. (1999), 'The state of macroeconomic forecasting', George Washington University, Center for Economic Research Discussion Paper No. 99-04

[7]     Zarnowitz, V. and Braun, P. (1992), 'Twenty-two years of the NBER-ASA Quarterly Outlook Surveys: aspects and comparisons of forecasting performance', NBER Working Paper 3965

[8]      Mellis, C. and Whittaker, R.  (1998), 'The Treasury forecasting record: some new results', *National Institute Economic Review*, no. 164, pp.65-79

[9]     Öller, L-E. and Barot, B., (1999), 'Comparing the accuracy of European GDP forecasts', National Institute of Economic Research, Stockholm, Sweden

[10]     M. L. Mehta, *Random Matrices*, Academic Press, Boston (1991)

[11]     Mullin, T. (1993) 'A dynamical systems approach to time series analysis', in T.Mullin, ed., *The Nature of Chaos*, Oxford Scientific Publications

[12]     Burns,A.F. and Mitchell,W.C. (1946), *Measuring the Business Cycle*, NBER

[13]     Cogley,T. and Nason,J.M. (1995), 'Output dynamics in real business cycle models', *American Economic Review*, vol.85, no.3, pp.492-511

[14]     Maddison, A. (1995), *Monitoring the World Economy 1820-1992*, OECD, Paris

[15]     Tong, H. (1990), *Non-linear Time Series: A Dynamic Systems Approach*, Oxford Statistical Science Series, Clarendon Press, Oxford

**Figure Captions**

**Figure 1 :**   Density of eigenvalues of the correlation matrices formed from the delay matrices of GDP growth for 17 capitalist economies during the period 1871-1994 (upper graph).

Also shown is the density of eigenvalues for the correlation matrices of 5,000 random matrices (formed from un-correlated time series) of the same dimension as the correlation matrix of the empirical GDP data (lower graph).

**Figure 2 :**   Inverse participation ratio versus eigenvalue for the United States GDP growth.

**Figure 1**

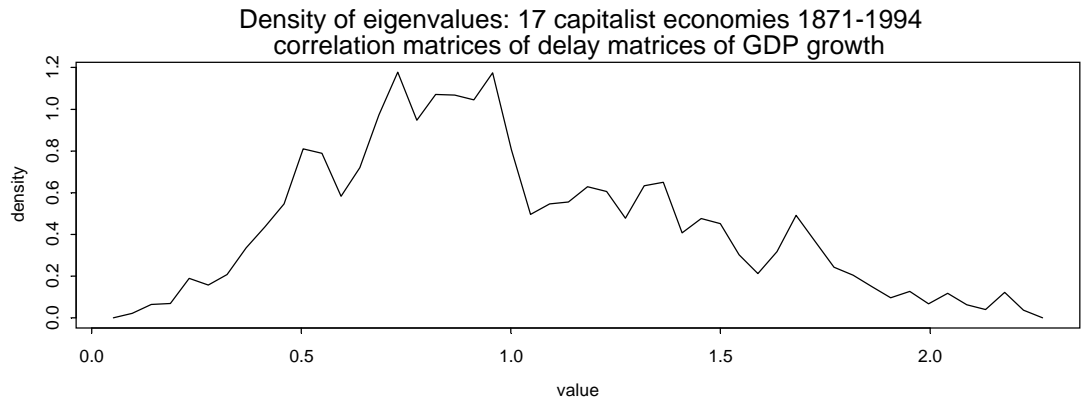

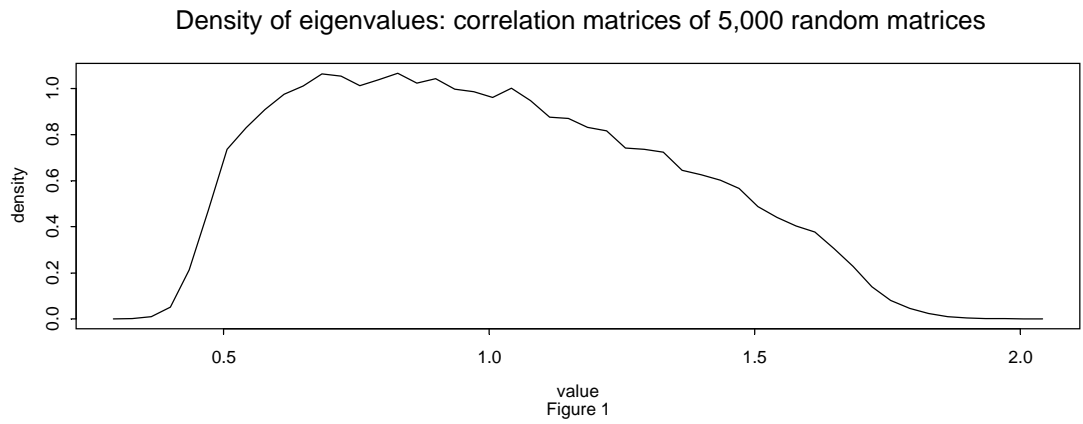

Figure 1

**Figure 2**

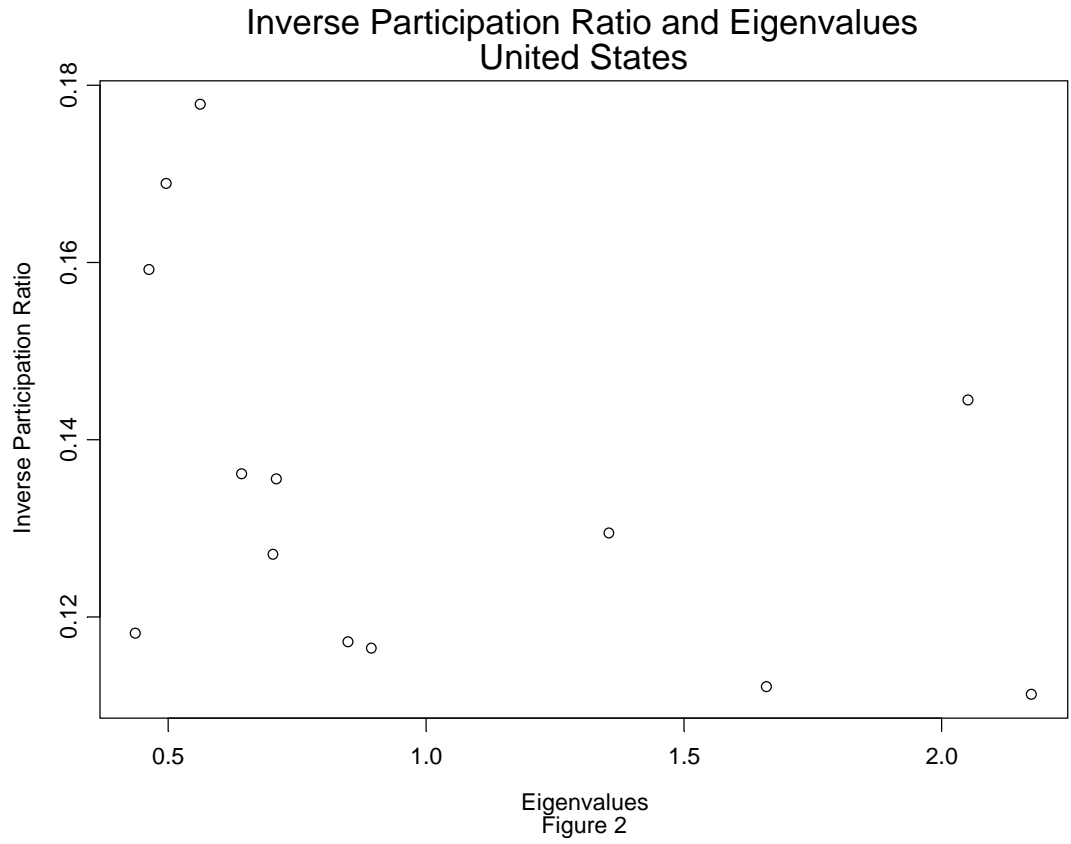
Figure 2